\documentclass[twocolumn,groupedaddress,superscriptaddress,showpacs,prb,letterpaper]{revtex4}

\usepackage{times,amsmath,amssymb}      
\usepackage{epsfig}
\usepackage{graphicx}% Include figure files
\usepackage{dcolumn}% Align table columns on decimal point
\usepackage{bm}% bold math

\newcommand{\vk}{{\bf k}}

\newcommand{\vf}{v_{\rm F}}

\newcommand{\lB}{\ell_{\rm B}}

\newcommand{\tp}{t_\perp}
\newcommand{\pk}{\phi_{\bf k}}
\newcommand{\mk}{| {\bf k} |}
\newcommand{\ek}{e^{i \pk}}
\newcommand{\emk}{e^{-i \pk}}
\newcommand{\epk}{\epsilon_{\bf k}}

\newcommand{\GA}{G_A ( \mk , \omega )}
\newcommand{\GB}{G_B ( \mk , \omega )}
\newcommand{\GAn}{G_A^n ( \mk , \omega )}
\newcommand{\GAnp}{G_A^{n+1} ( \mk , \omega )}
\newcommand{\GAnm}{G_A^{n-1} ( \mk , \omega )}
\newcommand{\GBn}{G_B^n ( \mk , \omega )}

\begin{document}

\title{Electronic states and Landau levels in graphene stacks}

\author{F. Guinea}
\affiliation{Instituto de  Ciencia de Materiales de Madrid, CSIC,
 Cantoblanco E28049 Madrid, Spain}

\author{A.~H. Castro Neto}
\affiliation{Department of Physics, Boston University, 590 
Commonwealth Avenue, Boston, MA 02215, USA}

\author{N.~M.~R. Peres}
\affiliation{Center of Physics and Departamento de F{\'\i}sica,
Universidade do Minho, P-4710-057, Braga, Portugal}

\begin{abstract}
We analyze, within a minimal model that allows analytical calculations, 
the electronic structure and Landau levels of graphene multi-layers with 
different stacking orders. We find, among other results, that electrostatic 
effects can induce a strongly divergent density of states in bi- and
tri-layers, reminiscent of one-dimensional systems. The density of states at 
the surface of semi-infinite stacks, on the other hand, may vanish at low 
energies, or show a band of surface states, depending on the stacking order.
\end{abstract}
%\pacs{73.20.-r; 73.20.Hb; 73.23.-b; 73.43.-f}


\maketitle

\section{Introduction}
\label{introduction}

Recent experiments show that single layer graphene, and stacks of graphene 
layers, have unusual electronic properties \cite{Netal04,Zetal05}, that may 
be useful in the design of new electronic devices
\cite{Betal04,Betal05,Netal05b}. Among them, we can mention the Dirac-like 
dispersion relation of a single graphene layer \cite{Netal05,Zetal05b}, the 
chiral parabolic bands in bilayers that lead to a new type of Quantum Hall 
effect \cite{Ketal06}, or the possibility of confining charge to the surface 
in systems with a few graphene layers \cite{Metal05c}. While the study of 
graphene multilayers is still in its infancy \cite{JCGN06}, the current 
experimental techniques allow for the extraction/production of multi-layers 
with the accuracy of a single atomic layer. The ability of creating stacks 
of graphene layers can provide an extra dimension to be explored in terms 
of electronic properties with functionalities that cannot be obtained with 
other materials.  

The nature of the stacking order in graphene multilayers has its origins in
the single graphene plane, a two-dimensional (2D) honeycomb lattice with two
inequivalent sub-lattices, A and B (two atoms per 2D unit cell). The
staggered stacking occurs in highly oriented pyrolytic graphite (HOPG) where
the graphene layers are arranged so that there two graphene layers per
three-dimensional (3D) unit cell in a $ABAB \cdots$ sequence (see
Fig.~\ref{lattice}). In this case, one of the atoms in one of the planes (say
B) is exactly in the center of the hexagon in the other plane. Hence, only
one of the sublattices (A) is on the top of the same sublattice in the other
layer. Nevertheless, this is not the only possible stacking observed in these 
systems \cite{BCP88}. Rhombohedral graphite with stacking sequence $ABCABC
\cdots$ has three layers per 3D unit cell (in the third layer the sublattice
B becomes on top of sublattice A of the second layer). Furthermore, stacking 
defects have been repeatedly observed in natural graphitic samples
\cite{B50,G67}, and also in epitaxially grown graphene films \cite{Retal05}. 
The richness of possible stacking configurations is due to the weak van der 
Waals forces that keep the layers together. Furthermore, the electronic
structure of the conduction band in graphene and bulk graphite is well 
described by a tight binding model which includes hopping between the 
$\pi$ orbitals in Carbon atoms, neglecting the remaining atomic orbitals 
which give rise to the $\sigma$ bands in graphite 
\cite{W47,M57,BN58,SW58,M64,DSM77,BCP88}. It is known that the low energy 
electronic structure depends on the stacking order in bulk samples 
\cite{H58,M69,TL88,CMGV91,CML92,CGM94}.  

In the following, we analyze the electronic structure and Landau levels of
graphitic structure with different stacking orders. We use a minimal tight
binding model that only includes interlayer hopping between nearest neighbor 
Carbon atoms in contiguous layers, and the ${\bf  k} \cdot {\bf p}$ expansion 
for the dependence on the momentum parallel to the layers. Within the 
${\bf   k} \cdot {\bf p}$ approximation, the two inequivalent corners 
of the Brillouin zone can be studied separately, and we will analyze only 
one of them. We do not consider the effects of disorder, already discussed 
in other publications \cite{PGN06,JCGN06}, but focus instead on the
analytical solution of the model that provide information on the unique 
properties of these systems.

The model used here does not take into account hopping which may be relevant
in order to describe the fine details of the band structure, such as the
trigonal distortions, that are also difficult to estimate using more
demanding local density functional (DFT) methods. Moreover, recent angle 
resolved photoemission experiments in crystalline graphite \cite{lanzara_new} 
show that those effects are very small. 
On the other hand, the methods used here, based on a mapping onto a set of
simple nearest-neighbor one-dimensional (1D) tight binding Hamiltonians is
quite feasible, allowing us to study a variety of situations, with and
without an applied magnetic field. It is worth 
noting that when the energy scales associated to the electron-electron
interaction and/or to in plane disorder are larger than the hopping
neglected here, the description presented below will be a reasonable ansatz
for the calculation of the effects of the electron-electron interaction. 

The model used is described in the Section \ref{model}, as well as a simple
scheme that allows the mapping of the problem of an arbitrary number of coupled 
graphene layers onto a 1D  tight binding model with nearest
neighbor hopping only. Section \ref{stacks} describes the main properties 
of few layer systems, also in the presence of a magnetic field. 
In Section \ref{semi} we discuss the bulk and surface electronic structure 
of semi-infinite stacks of graphene layers, with different stacking order. 
Section \ref{conclusions} contains the main conclusions of our work.

\section{The model.}
\label{model}

\begin{figure}
\begin{center}
\includegraphics*[width=5cm]{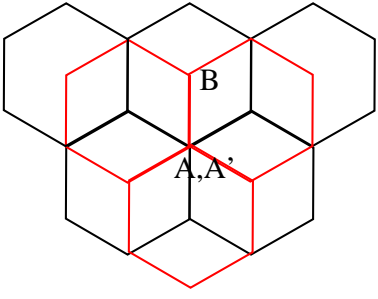}
\end{center}
\caption{Top view of the arrangement of the Carbon atoms in 
two neighboring graphene planes in the staggered stacking.}
\label{lattice}
\end{figure}

We consider the staggered stacking, where the layer sequence can be written as $1212 \cdots$,  and rhombohedral stacking, $123123 \cdots$ and combinations of the two.  We  do not consider hexagonal stacking, where all atoms in a given plane are on top of atoms in the neighboring layers, $111 \cdots$. In all stacking considered, the hopping between a pair neighboring layers takes place through half of the atoms in each plane, as schematically shown in Fig.~\ref{lattice}.

We describe the in-plane electronic properties of graphene using a low
energy, long wavelength, expansion around the $K$ and $K'$
corners of the 2D Brillouin Zone. This description 
uses as only input parameter the Fermi velocity, $\vf = 3 t a/2$, where 
$t$ ($\approx 3$ eV) is the hopping between $\pi$ orbitals located at nearest 
neighbor atoms, and $a = 1.42$\AA \, is the distance between Carbon atoms. 
We assume that hopping between planes takes place only between atoms which 
are on top of each other in neighboring layers. We denote the hopping
integral as $\tp$ ($\approx 0.1 t$). The whole model is defined 
by the parameters $\vf$ and $\tp$.  For instance, for the graphene bilayer 
(the 3D unit cell of the staggered stacking) the Hamiltonian has the form:  
$H = \sum_{{\bf k}} \Psi^{\dag} ({\bf k}) H_0({\bf k}) \Psi({\bf k})$, where ${\vk}=(k_x,k_y)$ is 
the 2D momentum measured relative to the $K$ ($K'$) 
point  (we use units such that $c=1=\hbar$), 
\begin{equation}
  \label{eq:Hkin0bilayer}
  H_0(\vk) = \vf
  \begin{pmatrix}
    0 & k e^{i \phi(\vk)} & \tp / \vf & 0 \\
    k e^{-i \phi(\vk)} & 0 & 0 & 0 \\
    \tp / \vf  & 0 & 0 & k e^{-i \phi(\vk)} \\
    0 & 0 & k e^{i \phi(\vk)} & 0
  \end{pmatrix},
\end{equation}
$\Psi^{\dag}({\bf k}) = \bigl( c^{\dag}_{\text{A},1,{\bf k}} \;
c^{\dag}_{\text{B},1,{\bf k}} \;
c^{\dag}_{\text{A},2,{\bf k}} \; c^{\dag}_{\text{B},2,{\bf k}} \bigr)$ is the electron 
spinor creation operator, and $\phi({\bf k}) =
\tan^{-1}(k_{y}/k_{x})$ is the 2D angle in momentum space.

In the case of the staggered stacking with $2N$ planes, the full Hamiltonian consists of 
$2N \times 2N$ matrix with $N/2$ blocks with size $4 \times 4$ given by 
(\ref{eq:Hkin0bilayer}).  The spinor operator is given by creation and annihilation 
operators for electrons in different planes: $c^{\dag}_{\alpha,n,{\bf k}}$ where $\alpha=A,B$
labels the sublattices in each plane, and $n=1, \cdots, N$ labels the
planes. We define the Green's functions:
\begin{eqnarray}
G^n_{\alpha,\beta}(\vk,t) = -i \langle {\cal T} c_{\alpha,n,\vk}(t) c^{\dag}_{\beta,n,\vk}(0) \rangle \, ,
\label{greenfunc}
\end{eqnarray}
where ${\cal T}$ is the time ordering operator, and its Fourier transform, $G^n_{\alpha,\beta}(\vk,\omega)$, that can be used to calculate the properties of these systems.

\begin{figure}
\begin{center}
\includegraphics*[width=8cm]{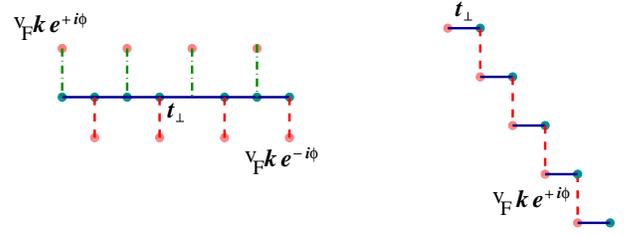}
\end{center}
\caption{Schematic view of the mapping of the problem of a stack of graphene
  layers onto a set of one dimensional problems. The hopping between sites
  are $\tp$ and $k e^{i \pm \phi}$, where $\phi = \arctan(k_y/k_x)$.
 Left: staggered stacking. Right:
  Rhombohedral stacking. Note that the phases $\phi$ can
  be gauged away in both cases.}
\label{TB_1D}
\end{figure}

It is easy to see from (\ref{eq:Hkin0bilayer}) that  $\GAn$
($\alpha=\beta=\text{A})$ and $\GBn$ ($\alpha=\beta=\text{B}$), satisfy the equations:
\begin{eqnarray}
\omega G_A^{2n} ( \mk , \omega ) &= &t_\perp [ G_A^{2n-1} ( \mk , \omega )
 +  G_A^{2n+1} ( \mk , \omega )] 
 \nonumber \\
&+& \vf \mk \ek G_B^{2n} ( \mk , \omega ) \, ,
 \nonumber \\
\omega G_B ^{2n} ( \mk , \omega ) &= &\vf \mk \emk G_A^{2n} ( \mk , \omega ) \, ,
 \nonumber \\ 
\omega G_A^{2n+1} ( \mk , \omega ) &= &t_\perp [ G_A^{2n} ( \mk , \omega ) + 
G_A^{2n+2} ( \mk , \omega )] 
\nonumber \\
&+& \vf \mk \emk G_B^{2n+1} ( \mk , \omega ) \, ,
\nonumber \\
\omega G_B^{2n+1} ( \mk , \omega ) &= &\vf \mk \ek G_A^{2n+1} ( \mk , \omega ) \, .
\label{g_HOPG}
\end{eqnarray}
The phase factors $\ek , \emk$ can be
gauged away by a gauge transformation, leading to a set of equations where the 
odd and even numbered layers are indistinguishable. We can integrate out the
$B$  sites in eq.(\ref{g_HOPG}),  and write an effective equation for $\GA$:
\begin{equation}
\left[ \omega - \frac{( \vf \mk )^2}{\omega} \right] \GAn = t_\perp [ \GAnp + \GAnm
] \, ,
\end{equation}
which is formally equivalent to the tight binding equations which describe a
1D chain with hopping energy $t_{\perp}$ and effective on site energy $( \vf \mk)^2/\omega$. The Green's function for the $B$ atoms can be obtained from $\GA$ 
using the expression:
\begin{equation}
%\left( \begin{array}{cc} \GAn & \GABn \\ \GABn & \GBn \end{array} \right)
\left[G^n_{\alpha,\beta}(\mk,\omega)\right]
\equiv \left( \begin{array}{cc} {\GAn}^{-1} + \frac{( \vf \mk )^2}{\omega} &
    \vf \mk \\ \vf \mk & \omega \end{array} \right)^{-1} \, ,
\end{equation}
so that, 
\begin{equation}
\GBn = \frac{1}{\omega} + \frac{( \vf \mk )^2}{\omega^2} \GAn \, .
\end{equation}  
A sketch of the resulting 1D tight binding model is given in Fig.~\ref{TB_1D}.

In the case of the rhombohedral stacking ($123123 \cdots$), 
a suitable gauge transformation can be used in order to make all layers equivalent,
although the two sublattices $A$ and $B$ within each layer remain different, 
leading to the following set of equations: 
\begin{eqnarray}
\omega G^n_A ( \omega ) &= &t_\perp G^{n-1}_B ( \omega ) + \vf \mk G^n_B (
\omega ) \, ,
\nonumber \\
\omega G^n_B ( \omega ) &= &t_\perp G^{n+1}_A ( \omega ) + \vf \mk G^n_A (
\omega ) \, ,
\label{g_turbo}
\end{eqnarray}
that determine the properties of the rhombohedral stacking. 

In a finite magnetic field $B$ the above equations can be modified with the
use of a Peierls substitution $\vk \to \vk + e {\bf A}$, where ${\bf A}$
is the vector potential ($\nabla \times {\bf A} = {\bf B}$). In this case,
the system breaks into Landau levels that, for a single graphene layer,
have energy \cite{PGN06}: 
\begin{eqnarray}
E(s,j) = s \omega_c \sqrt{j + 1} \, ,
\end{eqnarray} 
where $j = 0,1, \cdots$ labels the Landau levels, 
$s=\pm 1$ labels the electron and hole bands, 
\begin{eqnarray}
\omega_c = \sqrt{2} v_F/\lB = v_F \sqrt{2 e B} \, , 
\label{wc}
\end{eqnarray}  
is the cyclotron frequency, and $l_B$ is the cyclotron length. 

In the presence of magnetic field we can replace the momentum
label $\vk$ by the Landau label $j$ ($\mk \to \sqrt{j}/\lB$)
and the extension of the equations (\ref{g_HOPG}) become:
\begin{eqnarray}
\omega G_A^{2n} ( j,\omega ) = t_\perp [ G_A^{2n-1} (j,\omega )
 +  G_A^{2n+1} (j,\omega )] 
 \nonumber \\
+ \frac{\vf \sqrt{m}}{\lB}  G_B^{2n} (j-1,\omega) \, ,
 \nonumber \\
\omega G_B ^{2n} (j-1,\omega ) = \frac{\vf \sqrt{m}}{\lB} G_A^{2n} (j,\omega) \, ,
 \nonumber \\ 
\omega G_A^{2n+1} (j,\omega ) = t_\perp [ G_A^{2n} ( m , \omega ) + 
G_A^{2n+2} ( m , \omega )] 
\nonumber \\
+ \frac{\vf \sqrt{j+1}}{\lB} G_B^{2n+1} (j+1,\omega) \, ,
\nonumber \\
\omega G_B^{2n+1} (j+1,\omega ) =  \frac{\vf \sqrt{j+1}}{\lB} 
G_A^{2n+1} (j,\omega) \, .
\label{g_HOPG_B}
\end{eqnarray}
Note that, in this case, the two inequivalent layers in the unit cell cannot 
be made equivalent by a gauge transformation as in (\ref{g_HOPG}).

In rhombohedral case, the presence of a magnetic field, (\ref{g_turbo}) become:
\begin{eqnarray}
\omega G^n_A ( \omega ) &= &t_\perp G^{n-1}_B ( \omega ) + \frac{\vf}{\lB} \sqrt{j} G^n_B (\omega ) \, ,
\nonumber \\
\omega G^n_B ( \omega ) &= &t_\perp G^{n+1}_A ( \omega ) + \frac{\vf}{\lB}
\sqrt{j+1}  G^n_A ( \omega ) \, .
\label{g_turbo_B}
\end{eqnarray}

\section{Stacks of a few graphene layers.}
\label{stacks}
\subsection{Generic energy bands.}
Using the 1D tight binding mapping discussed in 
the preceding section, we can write an implicit equation for the
eigenenergies of a system with $N$ layers with the staggered stacking as:
\begin{equation}
\epsilon_{\vk} = \frac{( \vf \mk )^2}{\epsilon_{\vk}} + 2 t_\perp \cos \left(
  \frac{\pi n}{N+1} \right) \, ,
\end{equation}
where $n=1, \cdots N$, so that the dispersion becomes:
\begin{equation}
\epsilon_{\vk} \!=\! t_\perp \!\cos \left(\!\frac{\pi n}{N+1}\!\right)\!\pm\!
\sqrt{( \vf \mk )^2 \!+\! t_\perp^2 \!\cos^2 \left(\!\frac{\pi n}{N+1}\!\right)} \, .
\label{dispersion_ML}
\end{equation}

\subsection{Bilayer.}
\begin{figure}
\begin{center}
\includegraphics*[width=5cm]{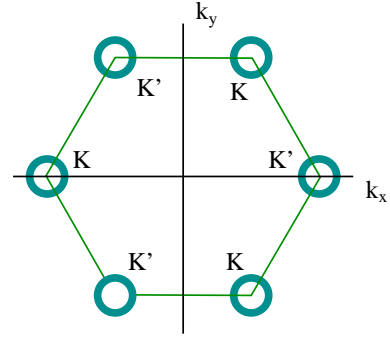}
\end{center}
\caption{The Fermi ring: the Fermi sea of a biased bilayer.}
\label{ring}
\end{figure}

A  number of properties of the present model applied to a graphene bilayer
can be found in ref.~[\onlinecite{MF06}]. 
We extend those results here to the case where
there is an electrostatic potential which makes the two layers inequivalent.
The Hamiltonian reads:
\begin{equation}
{\cal H}_{2L} ( \vk ) \!\!\equiv\!\! \left(\!\!\! \begin{array}{cccc} \Delta &\vf \mk \ek &\tp
    &0 \\ \vf \mk \emk &\Delta &0 &0 \\ \tp &0 &- \Delta &\vf \mk \emk \\ 0
    &0 &\vf \mk
    \ek &- \Delta \end{array} \!\!\!\right) \, ,
\label{hamil_2L}
\end{equation}
where the difference in the electrostatic  potentials in the two layers is $2
\Delta$. For $\vf \mk \ll \Delta \ll \tp$, the bands at low energy are given by
$\pm \epk$ where:
\begin{equation}
\epk \approx %\pm \sqrt{\Delta^2 - \frac{4 \Delta^2 \vf^2 \mk^2}{\tp^2} +
  %\frac{\vf^4 \mk^4}{\tp^2}} \approx 
 \Delta - \frac{\Delta \vf^2 \mk^2}{\tp^2} + \frac{\vf^4 \mk^4}{2 \Delta \tp^2} \, .
\end{equation}
Hence, the bands show an unusual ``mexican hat'' dispersion, with extrema at $k_0
\sim \Delta / \vf$ with energy $\epsilon_{vH} = \Delta - \Delta^3 / \tp^2$. 
Near these special points, the electronic density of states diverges as:
\begin{equation}
D_\pm ( \epsilon ) \propto \frac{\tp \sqrt{\Delta}}{\vf^2 \sqrt{\epsilon \mp
      \epsilon_{vH}}} \, ,
\label{dos_2L}
\end{equation}
which is the divergence seen in 1D systems. In order to understand the reason
for this 1D behavior consider the situation where the chemical potential is
above (or below) $\epsilon_{vH}$. In this case the Fermi surface looks like
a  ring (a Fermi ring), as shown in Fig.~\ref{ring}. At large radius $k_0$ the ring
approaches a 1D dispersion with nested Fermi surfaces, characteristic of 1D systems.

In the presence of a magnetic field, the Hamiltonian becomes:
\begin{equation}
{\cal H}_{2L} (j) \!\!\equiv\!\! \left(\!\! \begin{array}{cccc} \Delta &\frac{\vf
    \sqrt{j}}{\lB} &\tp
    &0 \\ \frac{\vf \sqrt{j}}{\lB} &\Delta &0 &0 \\ \tp &0 &- \Delta
    &\frac{\vf \sqrt{j+1}}{\lB}  \\ 0
    &0 &\frac{\vf \sqrt{j+1}}{\lB} &- \Delta \end{array} \!\!\right) \, ,
\label{hamil_2L_B}
\end{equation}
where $j$ is the index of the Landau level. For values of $j$ such that $(
\vf \sqrt{j} ) / \lB \ll \Delta \ll \tp$, we obtain:
\begin{equation}
\epsilon ( j ) \!\!=\!\! - \frac{\vf^2 \Delta}{\lB^2 \tp^2} \left(\!j\!+\!\frac{1}{2}\!
\right)
\!\pm\! \sqrt{\Delta^2 \left(\!1\!-\!\frac{\vf^2}{2 \lB^2 \tp^2}\!\right)^2 
\!+\! \frac{\vf^4 j ( j+1 )}{\lB^4 \tp^2}} \, .
\end{equation}

\subsection{Trilayer.}
\subsubsection{HOPG stacking (ABA).}
The model used here leads to six energy bands in a trilayer. In the absence
of electrostatic potentials that break the equivalence of the three layers,
we can use eq.(\ref{dispersion_ML}), to obtain:
\begin{eqnarray}
\epk &= & \pm \vf \mk \, , 
\nonumber 
\\
\epk &= &\frac{\tp \sqrt{2}}{2} \pm \sqrt{\frac{\tp^2}{2} + \vf^2 \mk^2} \, .
\end{eqnarray}
There is a band with Dirac-like linear dispersion, and two more bands which
disperse quadratically near $\epsilon = 0$, as in a bilayer. The Landau
levels, at low energies are:
\begin{eqnarray}
\epsilon ( j ) &= &\pm \frac{\vf \sqrt{j}}{\lB} \, , \nonumber \\
\epsilon ( j ) &= &\pm \frac{\vf^2 \sqrt{j (j+1 )}}{\lB^2 \tp}
\end{eqnarray}
Hence, the spectrum can be viewed as a superposition of the corresponding
spectrum for the Dirac equation, and that obtained for a bilayer.

\begin{figure}
\begin{center}
\includegraphics*[width=6cm]{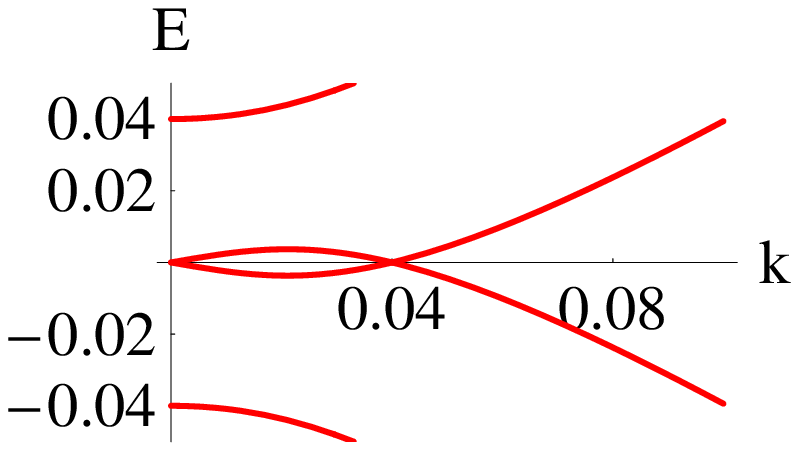}
\includegraphics*[width=6cm]{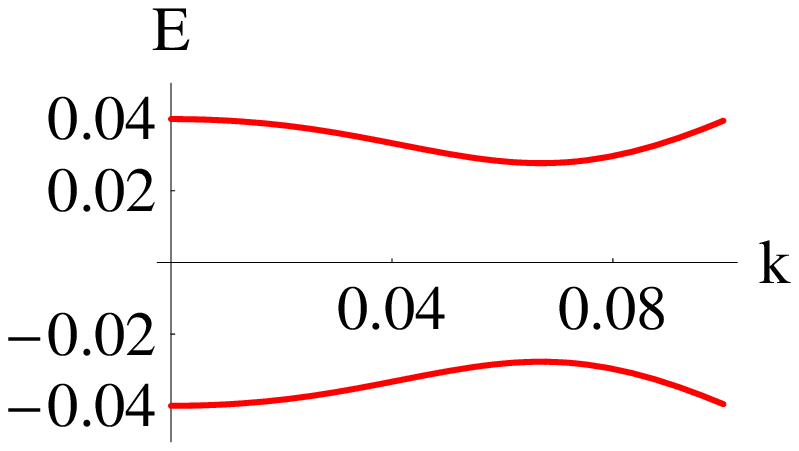}
\end{center}
\caption{Low energy bands of a trilayer with an electrostatic field which
  hreaks the equivalence of the three layers (see text). Top: HOPG stacking,
  $ABA$. Bottom: rhombohedral stacking, $ABC$. The parameters used
  are $\vf = 1, \tp = 0.1 , \Delta = 0.04$.}
\label{3L}
\end{figure}

The calculation of the energy levels in the presence of electrostatic fields
which break the symmetry between the three layers is more complex, and cannot
be performed fully analytically. We consider the case when the upper layer is
at potential $\Delta$, the lower layer is at potential $- \Delta$, and the
middle layer is at zero potential. The Hamiltonian is:
\begin{widetext}
\begin{equation}
{\cal H}_{3L} ( \vk )\!\!=\!\! \left(\!\!\begin{array}{cccccc} \Delta &\vf \mk \ek
    &0 &0 &0 &0 \\ \vf \mk \emk &\Delta &\tp &0 &0 &0 \\ 0 &\tp &0 &\vf \mk
    \emk &\tp &0 \\ 0 &0 &\vf \mk \ek &0 &0 &0 \\ 0 &0 &\tp &0 &- \Delta
    &\vf \mk \ek \\ 0 &0 &0 &0 &\vf \mk \emk &- \Delta \end{array}\!\!\right)
    \, ,
\end{equation}
\end{widetext}
In the limit $\vf \mk \ll \Delta \ll \tp$, one can use perturbation theory to
obtain:
\begin{equation}
\epk \approx \frac{\Delta \vf \mk}{\sqrt{2} \tp} \left( 1 - \frac{\vf^2
    \mk^2}{\Delta^2} \right) \, .
\end{equation}
This equation, although approximate, describes correctly the existence of
states at zero energy when $\vf \mk = \Delta$. The dispersion shows extrema
at momenta $k_0 \propto \Delta / \vf$ with energy $\epsilon_{vH} \propto  \Delta^2 /
\tp$, leading to a divergent density of states:
\begin{equation}
D_\pm ( \epsilon ) \propto \frac{\Delta \sqrt{\tp}}{\vf^2 \sqrt{\epsilon \mp
      \epsilon_{vH}}}
\label{dos_3L}
\end{equation} 
This expression, as in the case of eq.~(\ref{dos_2L}), is typical of 1D systems. 
The low energy bands of a trilayer with $ABA$ stacking 
are shown in the upper panel of Fig.[\ref{3L}].

\subsubsection{Rhombohedral stacking (ABC).}
The low energy bands of a trilayer with the $ABC$ stacking differ
significantly from those for the $ABA$ stacking discussed previously. The $6
\times 6$ hamiltonian is:
\begin{widetext}
\begin{equation}
{\cal H}_{3L} ( \vk ) \equiv \left( \begin{array}{cccccc} \Delta &\vf \mk \ek
    &0 &0 &0 &0 \\ \vf \mk \emk &\Delta &\tp &0 &0 &0 \\ 0 &\tp &0 &\vf \mk
    \ek &0 &0 \\ 0 &0 &\vf \mk \emk &0 &\tp &0 \\ 0 &0 &0 &\tp &- \Delta
    &\vf \mk \ek \\ 0 &0 &0 &0 &\vf \mk \emk &- \Delta \end{array} \right)
\label{hamil_3L_eh}
\end{equation}
\end{widetext}
Assuming that $\Delta \ll \tp$, only two out of
the six $\pi$ bands lie at energies $| \epsilon | \ll \tp$. These
bands are derived from the orbitals at the $B$ sublattice in the two outermost
layers, given by the first and sixth entry in eq.(\ref{hamil_3L_eh}). One can
define an effective $2 \times 2$ hamiltonian:
\begin{equation}
{\cal H}_{3L \, \,  eff} \equiv \left( \begin{array}{cc} - \Delta 
+ \frac{\Delta \vf^2
      \mk^2}{ \tp^2} &\frac{\vf^3 \mk^3}{\tp^2} 
\\ \frac{\vf^3 \mk^3}{\tp^2} & \Delta - \frac{\Delta \vf^2
      \mk^2}{\tp^2} \end{array} \right) 
\end{equation}
The dispersion is cubic at low momenta, $\epk \approx ( \vf \mk )^3 /
\tp^2$. The density of states is $D ( \epsilon ) \propto \tp^{4/3} / (
\vf^{2} \epsilon^{1/3} )$. The low energ bands of a trilayer with $ABC$
stacking is shown in the lower panel of Fig.[\ref{3L}].

The effective $2 \times 2$ hamiltonian in the presence of an applied magnetic
field is, for $\Delta = 0$:
\begin{equation}
{\cal H}_{3L \, \, rhombo} \equiv \left( \begin{array}{cc} 0 &\frac{\vf^3
      \sqrt{n ( n+1 ) ( n+2 )}}{\lB^3 \tp^3} \\ \frac{\vf^3 \sqrt{n ( n+1 ) (
      n+2 )}}{\lB^3 \tp^3} &0 \end{array} \right)
\end{equation}
There are, in addition, states localized in the outermost layers, with Landau
index $n=0$, and in two layers, with Landau index $n=1$. It is interesting to
note that Landau levels associated to different $K$ points are localized in
different layers.
\section{Semi-infinite stacks of graphene layers.}
\label{semi}

In this section we consider the case of a graphene multilayer with a surface termination.
In the case of the staggered stacking, we can use eq.(\ref{g_HOPG}), and obtain for the Green's function in the bulk as:
\begin{equation}
\GA = \frac{\omega}{\sqrt{[ \omega^2 - ( \vf \mk )^2 ] - 4 t_\perp^2
    \omega^2}} \, .
\end{equation} 

\begin{figure}
\begin{center}
\includegraphics*[width=8cm]{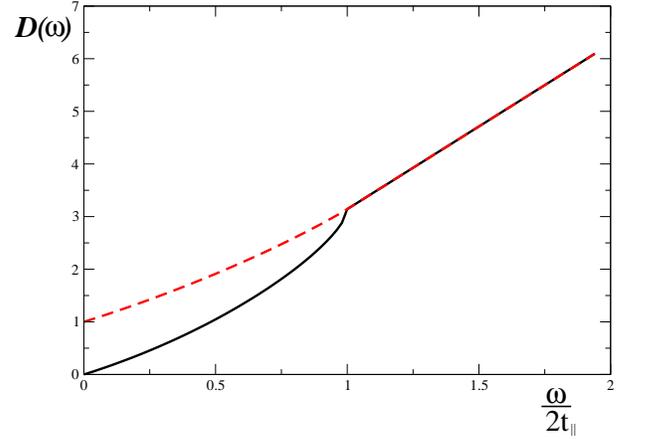}
\end{center}
\caption{Density of states for the staggered stacking. 
Continuous line: ${\rm Im} G_B ( \omega)$. Dashed line: ${\rm Im} G_A ( \omega)$. }
\label{DOS}
\end{figure}

We can obtain the local density of states by integrating $\GA$ and $\GB$ over
the parallel momentum $\vk$. This integral can be performed analytically, and
we obtain, for the bulk:
\begin{eqnarray}
G_A ( \omega ) \!&=&\!\frac{\omega}{\vf^2} \ln \left[ \frac{ ( \vf \Lambda )^2}{
    \omega ( \omega \!+\! \sqrt{\omega^2 \!-\! 4 t_\perp^2} )} \right] \, ,
    \nonumber \\
G_B ( \omega ) \!\!\!&= &\!\!\! \frac{\sqrt{\omega^2 \!-\! 4 t_\perp^2}}{\vf^2} \!+\!
    \frac{\omega}{\vf^2} \!
\ln \!\!\left[\! \frac{ ( \vf \Lambda )^2}{
    \omega ( \omega + \sqrt{\omega^2 - 4 t_\perp^2} )} \!\right] \! \!.
\label{green}
\end{eqnarray}
The local density of states is given by (see also ref.~[\onlinecite{JCGN06}]):
\begin{eqnarray}
{\rm Im} G_A ( \omega ) &= & \frac{\omega}{\vf^2} \left\{
\left[\frac{\pi}{2} + \arctan \left( \frac{\omega}{\sqrt{4 t_\perp^2 -
        \omega^2}} \right)\right] \right. 
\nonumber
\\
&\times& \left. \Theta(2 \tp-\omega) +  \pi \Theta(\omega-2 \tp)\right\} \, ,
\nonumber
\\
{\rm Im} G_B ( \omega )\!\!\!&= &\!\!\!\! \left\{\!\!\frac{\sqrt{4 t_\perp^2 -
    \omega^2}}{\vf^2} \!\!+\!\! \frac{\omega}{\vf^2} \left[ \frac{\pi}{2} \!+\! \arctan
    \left(\!\frac{\omega}{\sqrt{4 t_\perp^2 - \omega^2}} \right)\!\right]\!\right\} 
    \nonumber
    \\
 &\times& \Theta(2 \tp -\omega) + \frac{\pi}{\vf^2} \Theta(\omega-2 \tp) \, .
\end{eqnarray}
The density of states in the bulk of the staggered stacking is shown in 
Fig.~\ref{DOS}. 

\begin{figure}
\begin{center}
\includegraphics*[width=8cm]{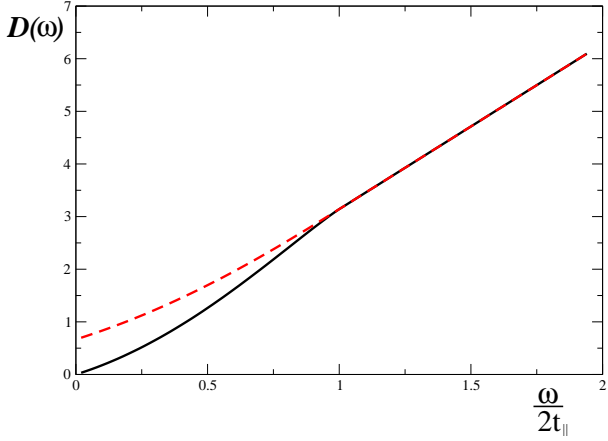}
\end{center}
\caption{Density of states at the surface layer. 
 Continuous line: ${\rm Im} G_B ( \omega)$. Dashed line: ${\rm Im} G_A ( \omega)$. }
\label{DOS_surface}
\end{figure}

We can use the equivalence to a 1D model for each value of
$\vk$, (\ref{g_HOPG}) and (\ref{g_turbo}), in order to obtain the Green's
function at the surface layer of a semi-infinite system:
\begin{eqnarray}
G^{\rm surface}_A ( \omega ) &= &\frac{\sqrt{[ \omega^2 - ( \vf \mk )^2 ] - 4 t^2 \omega^2}}{2 t^2
  \omega} - \frac{\omega}{2 t_\perp^2} \, , \nonumber \\
G^{\rm surface}_B ( \omega ) &= &\frac{1}{\omega} - \frac{\vf^2 \mk^2}{\omega^2} G^{\rm surface}_A ( \omega ) \, ,
\end{eqnarray}
for staggered stacking. The local density of states, obtained
after integrating these expressions over ${\vk}$ is shown in Fig.~\ref{DOS_surface}. 

The 1D tight binding model which describes the electronic bands
in rhombohedral graphite (stacking order $123123 \cdots$) is 
given in (\ref{g_turbo}).
Using this equation, the Green's function, integrated over the perpendicular momentum,
$k_\perp$, is given by:
\begin{equation}
G^n ( {\vk} , \omega ) \!\!=\!\! \frac{\omega}{\sqrt{\left[ \omega^2 \!-\! ( \vf \mk \!+\! \tp )^2
    \right] \left[ \omega^2 \!-\! ( \vf \mk \!+\! \tp )^2 \right]}} .
\end{equation}
The local density of states can be obtained by integrating this expression
over $\vk$:
\begin{equation}
{\rm Im} G^n ( \omega ) = \frac{| \omega |}{\vf^2} \, .
\label{g_local_turbo}
\end{equation}
The material, within this approximation, is a semi-metal, with vanishing
density of states at the Fermi level at half filling, $\omega = 0$. Note that
the density of states at low energies is independent of the value of $\tp$.

\begin{figure}
\begin{center}
\includegraphics*[width=7cm,angle=-90]{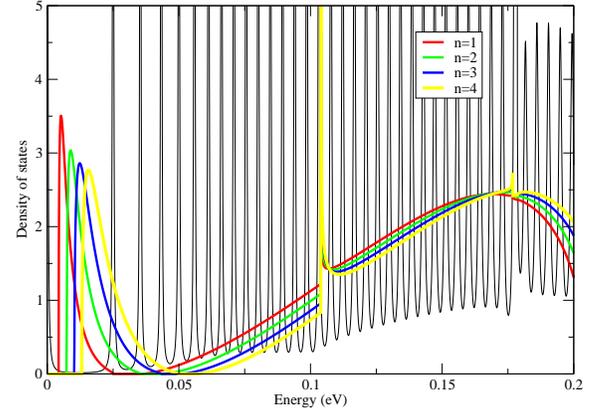}
\end{center}
\caption{Average density of states of a stack of 50 graphene layers with
  rhombohedral order, $ABCABC \cdots$, embedded into the staggered stacking, $ABAB
  \cdots$. The parameters used are: $\tp = 0.1$eV, and $B = 1$T ($\lB \approx
  25.7$nm, $\vf / \lB \approx 0.025$eV). The bands of some of the lowest 
Landau levels of the staggered stacking, calculated for the same parameters, are
  shown for comparison.}
\label{Landau_levels}
\end{figure}

In the presence of a magnetic field, we use (\ref{g_turbo_B}).
We can integrate out the Green's functions for the sites in a given
sublattice and obtain:
\begin{eqnarray}
\left( \omega - \frac{\tp^2}{\omega} - \frac{\vf^2 m}{\lB^2 \omega} \right)
G^n_A ( \omega ) = \frac{\tp \vf \sqrt{m}}{\lB \omega} G^{n-1}_A ( \omega )
\nonumber
\\
+ \frac{\tp \vf \sqrt{n+1}}{\lB \omega} G^{n+1}_A ( \omega ) \, .
\end{eqnarray}
These equations are formally equivalent to those obtained for the
wavefunctions of a displaced 1D harmonic oscillator:
\begin{equation}
{\cal H} \equiv \epsilon_0 + \omega_0 b^\dag b + g ( b^\dag + b ) \, ,
\label{h_eff}
\end{equation}
with the correspondence: $\epsilon_0 \leftrightarrow \tp^2 / \omega$, $\omega_0
\leftrightarrow \vf^2 / ( \lB^2 \omega )$,  $g \leftrightarrow ( \tp \vf ) / (
\lB \omega )$. The eigenenergies of (\ref{h_eff}) are $\epsilon_m =
\epsilon_0 - g^2 / \omega_0 + m \omega_0$, and we can write the eigenenergies associated to eq.(\ref{g_turbo_B}) as:
\begin{equation}
\epsilon_m = \frac{\tp^2}{\epsilon_m}  -
\frac{(\tp)^2}{\epsilon_m} + m \frac{\vf^2}{\lB^2 \epsilon_m} \, ,
\end{equation}
and, finally, we obtain:
\begin{equation}
\epsilon_m = \frac{\vf \sqrt{m}}{\lB} \, .
\end{equation}
The spectrum, in an applied magnetic field, is discrete, and equal to that in
graphene, as well as the local density of states in the absence of the field, 
eq.~(\ref{g_local_turbo}). The discreteness of the spectrum survives when a stack of rhombohedral graphene is embedded into the staggered stacking, as shown in
Fig.~\ref{Landau_levels}.

\begin{figure}
\begin{center}
\includegraphics*[width=8cm]{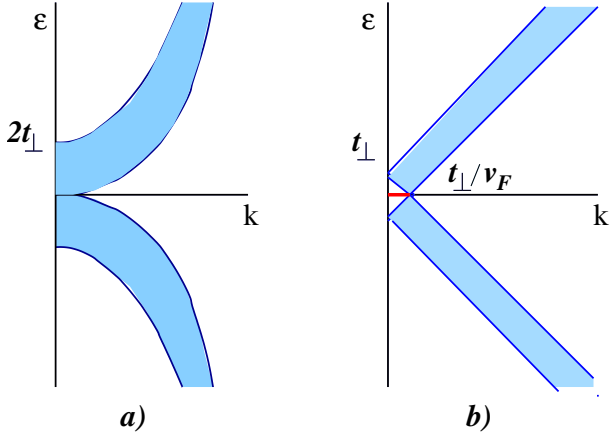}
\end{center}
\caption{Sketch of the projected density of states at the surface, 
as function of parallel momentum. Left: staggered stacking. 
Right: rhombohedral stacking (the line at $\epsilon=0$ stands for 
a dispersionless band of surface states).}
\label{dos_surface}
\end{figure}

Using the model described in Section \ref{model}, we obtain for the projected density
of states:
\begin{eqnarray}
G^{\rm surface}_A ( \omega ) &=& \frac{\omega^2 - \tp^2 - ( \vf \mk )^2}{2 \tp \vf \mk \omega}
\nonumber
\\
&-&\!\! \frac{\sqrt{\left[ \omega^2 \!-\! ( \vf \mk \!-\!
      \tp )^2 \right] \left[ \omega^2 \!-\! ( \vf \mk \!+\! \tp )^2 \right]}}{2 \tp
      \vf \mk \omega}   \, ,
      \nonumber 
      \\
G^{\rm surface}_B ( \omega ) &= & \frac{1}{\omega - ( \vf \mk )^2 G^{\rm surface}_A ( \omega )} \, .
\end{eqnarray}
The real part of these Green's functions has a pole at $\omega = 0$ for
$\mk \le \tp / \vf$, indicating the presence of a surface state. 
The projected density of states, as function of $\mk$, is sketched in
Fig.~\ref{dos_surface}. 

\begin{figure}
\begin{center}
\includegraphics*[width=6cm]{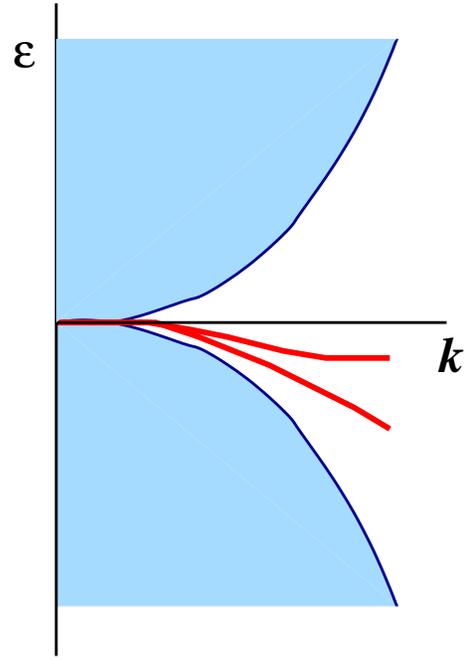}
\end{center}
\caption{Sketch of the projected density of states of a stack of graphene
  layers with ordering $CABAB \cdots$, in presence of a potential shift in
  the topmost layer. The lines outside the continuum of states 
stand for a two bands of of surface states.}
\label{surface_states}
\end{figure}

We can also analyze modifications of the surface, induced by electrostatic
potentials or local changes of the stacking order. A shift of the topmost
layer by a potential $\epsilon$ does not change qualitatively the projected
band structure shown in Fig.~\ref{dos_surface}, unless $\epsilon \ge
\tp$. On the other hand, a stacking of the type $1232323 \cdots$ leads to a
surface band, which can be shifted by an external potential, $\epsilon$,
applied to the topmost layer, labeled $1$. As the structure
looks like a perfect staggered stacking beyond this layers, and the couplings are
local, the Green's function in the topmost layer is given by:
\begin{equation}
\left[G_{\alpha,\beta}(\omega)\right] \equiv \left( \begin{array}{cc}
    \omega - \epsilon &\vf \mk \\ \vf \mk &\left( G_A^{{\rm stag.}} \right)^{-1} + \frac{( \vf
    \mk )^2}{\omega} \end{array} \right)^{-1}
\end{equation} 
where the labels $A , B$ correspond to the two inequivalent sites in the
topmost layer. A sketch of the resulting projected electronic structure is
shown in Fig.~\ref{surface_states}. 
This result is consistent with the observation of at
least two frequencies in the Shubnikov-de Haas experiments in graphene multilayers with
induced carriers at the surface reported in ref.~[\onlinecite{Metal05c}].

\section{Conclusions}
\label{conclusions}

We have analyzed a minimal model for the electronic structure of stacks of
graphene planes, which allows us to obtain a number of interesting results
analytically. In the following, we outline some of the most relevant ones:
\begin{itemize}
\item
Bilayers and trilayers, in the presence of electrostatic fields that break
the equivalence of the layers, develop van Hove singularities where the
density of states behaves in a quasi 1D fashion, $D ( \epsilon )
\propto ( \epsilon - \epsilon_{vH} )^{-1/2}$.
\item
The Landau levels of a trilayer, in the absence of electrostatic effects, are
given by a set of levels which depend on the field in the same way as for a
single layer, $\epsilon ( j ) = \pm ( \vf \sqrt{j} ) / \lB$, and another set
which is equivalent to the levels in a bilayer, $\epsilon ( j ) = \pm (
( \vf^2 \sqrt{j(j+1)} ) / \tp \lB^2 )$. A comparison of  the two sets will
allow for an independent measurement of the value of $\tp$.
\item
The surface density of states in staggered stacking vanishes at zero energy at the sites with a
nearest neighbor in the next layer. This result may explain the difference in the
images of the atoms at the two sublattices observed in STM measurements of
graphite surfaces\cite{TL88}.
\item
The bulk local density of states of rhombohedral stacking ($ABCABC \cdots$) 
vanishes at zero energy, and is independent of the value of the
interlayer hopping, $\tp$. The spectrum of bulk Landau levels is discrete,
unlike most three dimensional systems, and shows the same field dependence as
that of a single graphene layer, with independence of the value of
$\tp$. Hence, discrete, quasi-2D Landau levels may exist in
nominally 3D samples, helping to explain the experiments in
ref.~[\onlinecite{Eetal03b}].

\item
The surface of rhombohedral, and of staggered stacking with a rhombohedral
termination ($CABAB \cdots$) has surface states, with a well
defined dispersion as function of the parallel momentum. This result may
help to explain the observation of  Shubnikov-de Haas oscillations in graphitic
systems with a highly doped surface \cite{Metal05c}.
\end{itemize}

\section{Acknowledgments.}

We have benefited from interesting discussions on these topics with
C. Berger, P. Esquinazi, W. A. de Heer, A. K. Geim,
A. F. Morpurgo, J. Nilsson, K. Novoselov, J. G. Rodrigo, and S. Vieira.
A.H.C.N. was supported through NSF grant DMR-0343790. N. M. R. P. thanks 
ESF Science Programme INSTANS 2005-2010, and FCT under the grant POCTI/FIS/58133/2004.
F. G. acknowledges
funding from MEC (Spain) through grant FIS2005-05478-C02-01 and the European
Union Contract 12881 (NEST).

\bibliography{graphite0_15}
\end{document}